%% file: SurveyMobility.tex
\documentclass[twocolumn]{paper}

\usepackage[english]{babel}
\usepackage[T1]{fontenc}
\usepackage[utf8]{inputenc}

\usepackage{graphicx}                       % Include figure files

\usepackage{balance}

\usepackage{fullpage}

\usepackage{color}

\newcommand*{\affaddr}[1]{#1} % No op here. Customize it for different styles.
\newcommand*{\affmark}[1][*]{\textsuperscript{#1}}

\graphicspath{{./figures/}}

\begin{document}

\author{
\large Carlos Sarraute\affmark[1,*] and Martin Minnoni\affmark[1] \vspace{.05in}\\
\affaddr{\affmark[\scriptsize 1]\small Grandata Labs, San Francisco, CA, USA}\\
\affaddr{\affmark[\scriptsize *]\small Corresponding author: charles@grandata.com}\\
}

\title{Brief survey of Mobility Analyses based on Mobile Phone Datasets}

\date{}
\maketitle{}

\begin{abstract}
This is a brief survey of the research performed by Grandata Labs 
in collaboration with numerous academic groups around the world
on the topic of human mobility.
A driving theme in these projects is to use and improve Data Science techniques to understand mobility, as it can be observed through the lens of mobile phone datasets.
We describe applications of mobility analyses for urban planning, prediction of data traffic usage, building delay tolerant networks, generating epidemiologic risk maps and measuring the predictability of human mobility.

\end{abstract}

%-----------------------------------------------------
\section{Introduction}
%-----------------------------------------------------

The mission of Grandata's research lab is to improve our understanding of Human Dynamics through the analysis of massive datasets coming from mobile phone companies and other industries. This research has been performed in collaboration with numerous academic groups at MIT, INRIA and ENS Lyon, LNCC, UBA and many others.

We provide here a brief review of our research, intended to serve as an introduction and guideline to the research papers
on mobility aspects.

This brief survey focuses on the analysis of mobility in space, investigating the important locations of the users' trajectories, and how they can be used to infer their participation in large social events.
The study of mobility has numerous applications, such as 
urban planning (Section~\ref{urban-planning}),
data traffic usage prediction (Section~\ref{data-traffic-usage}),
building delay tolerant networks (Section~\ref{delay-tolerant-networks}),
epidemiology (Section~\ref{epidemiology}),
and predictability of human mobility (Section~\ref{mobility-predictability}).

\input{mobility}

%-----------------------------------------------------
% References
%-----------------------------------------------------
\bibliographystyle{ieeetr}
\balance
\bibliography{../GD_works}{}
%-----------------------------------------------------

\end{document}

%% file: mobility.tex
%!TEX root = SurveyMobility.tex

%-----------------------------------------------------
\section{Mobility and Urban Planning}
\label{urban-planning}
%-----------------------------------------------------

\begin{figure*}[th]
\begin{minipage}{.33\textwidth}
  \centering
  \includegraphics[width=0.95\textwidth]{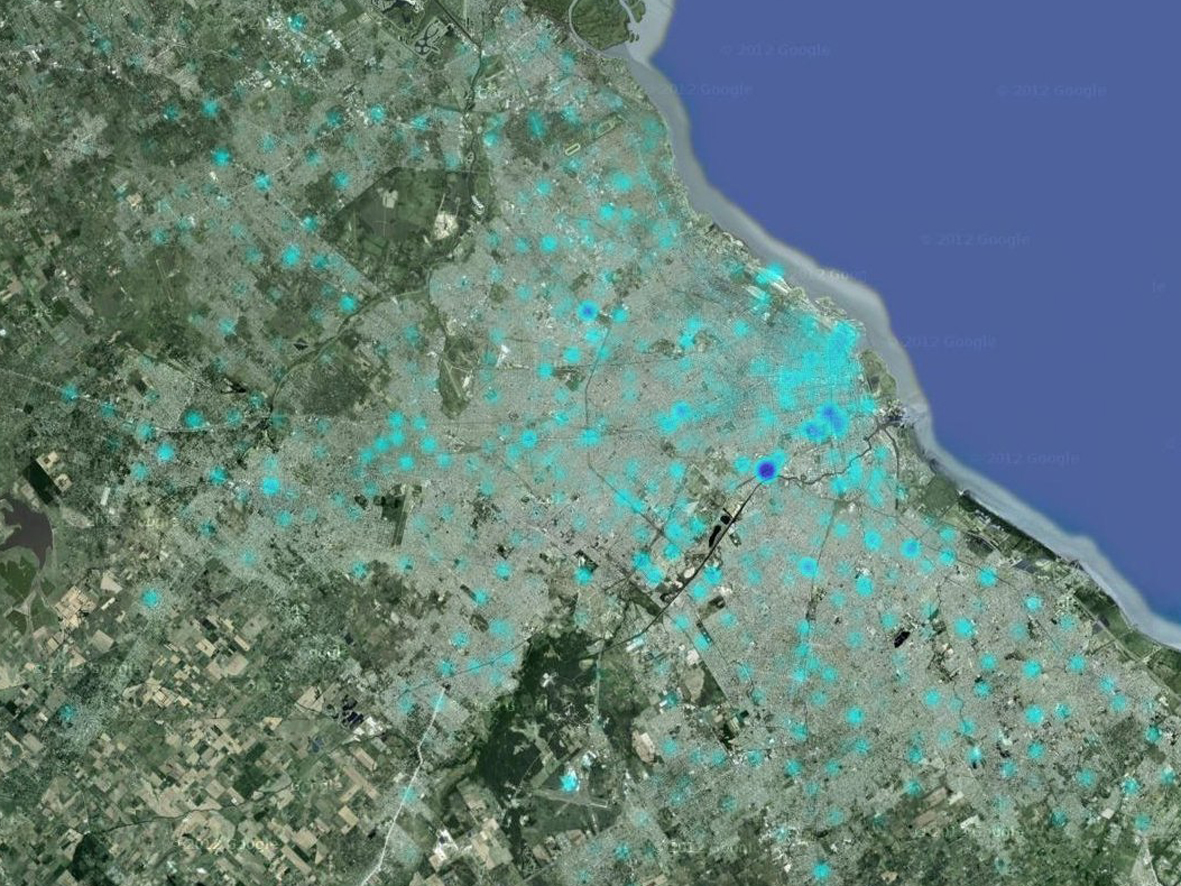}
  (a) 6 a.m.
\end{minipage}
%\hfill
\begin{minipage}{.33\textwidth}
  \centering
  \includegraphics[width=0.95\textwidth]{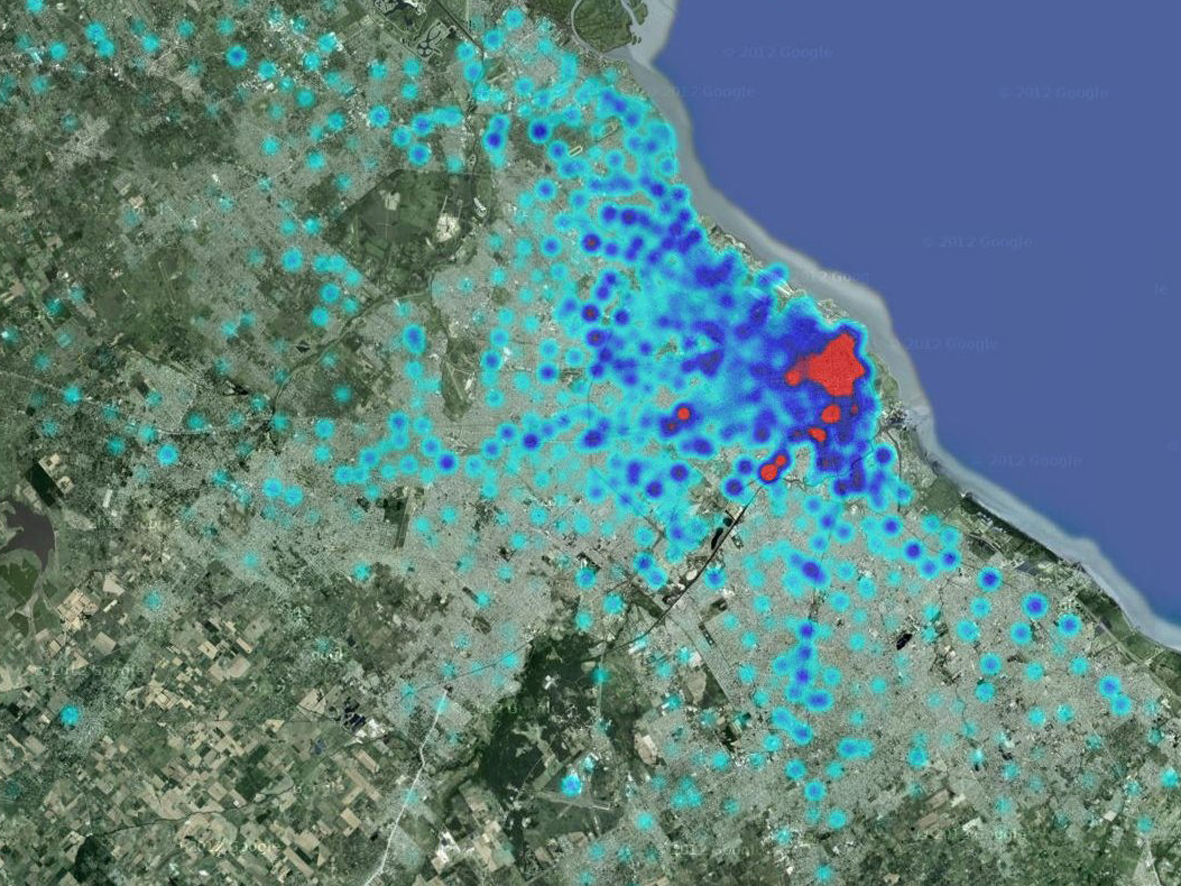}
  (b) 8 a.m.
\end{minipage}
\begin{minipage}{.33\textwidth}
  \centering
  \includegraphics[width=0.95\textwidth]{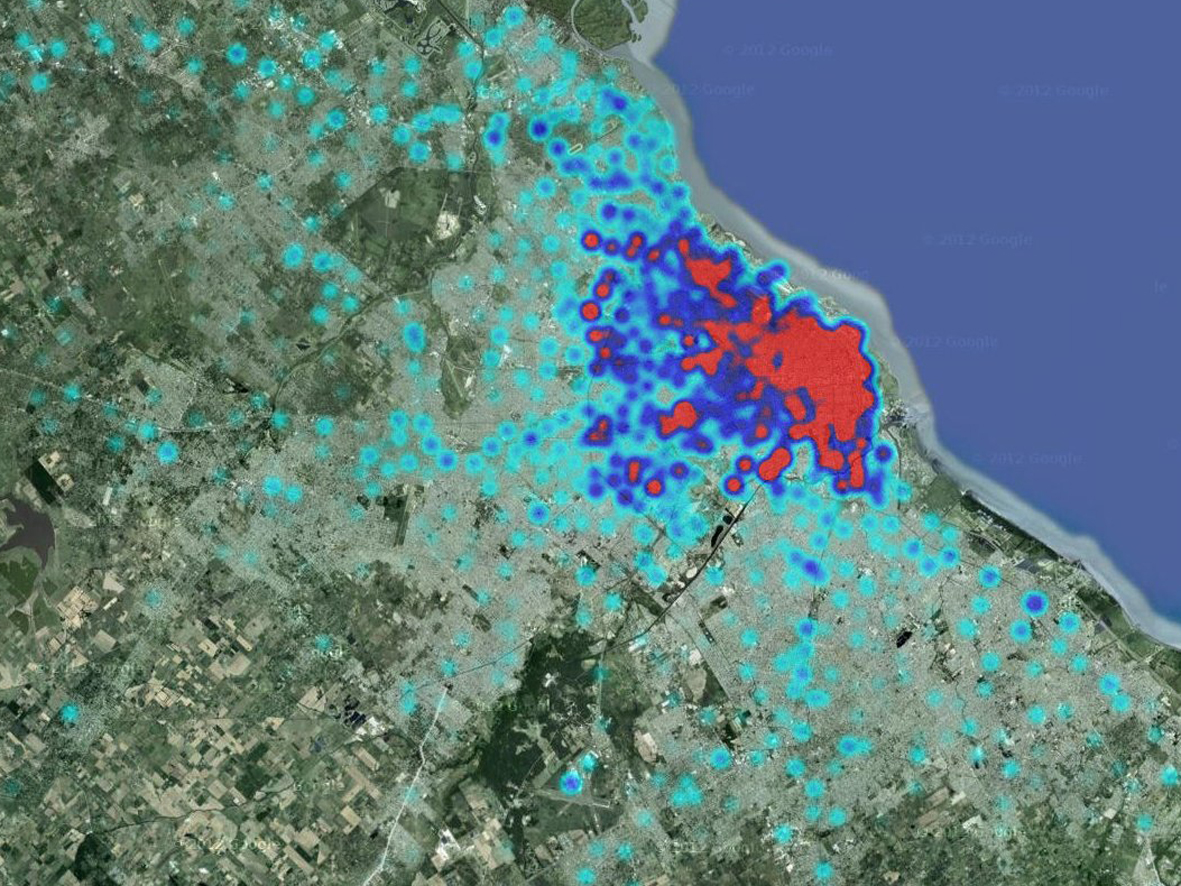}
  (c) 10 a.m.
\end{minipage}

\begin{minipage}{.33\textwidth}
  \centering
  \includegraphics[width=0.95\textwidth]{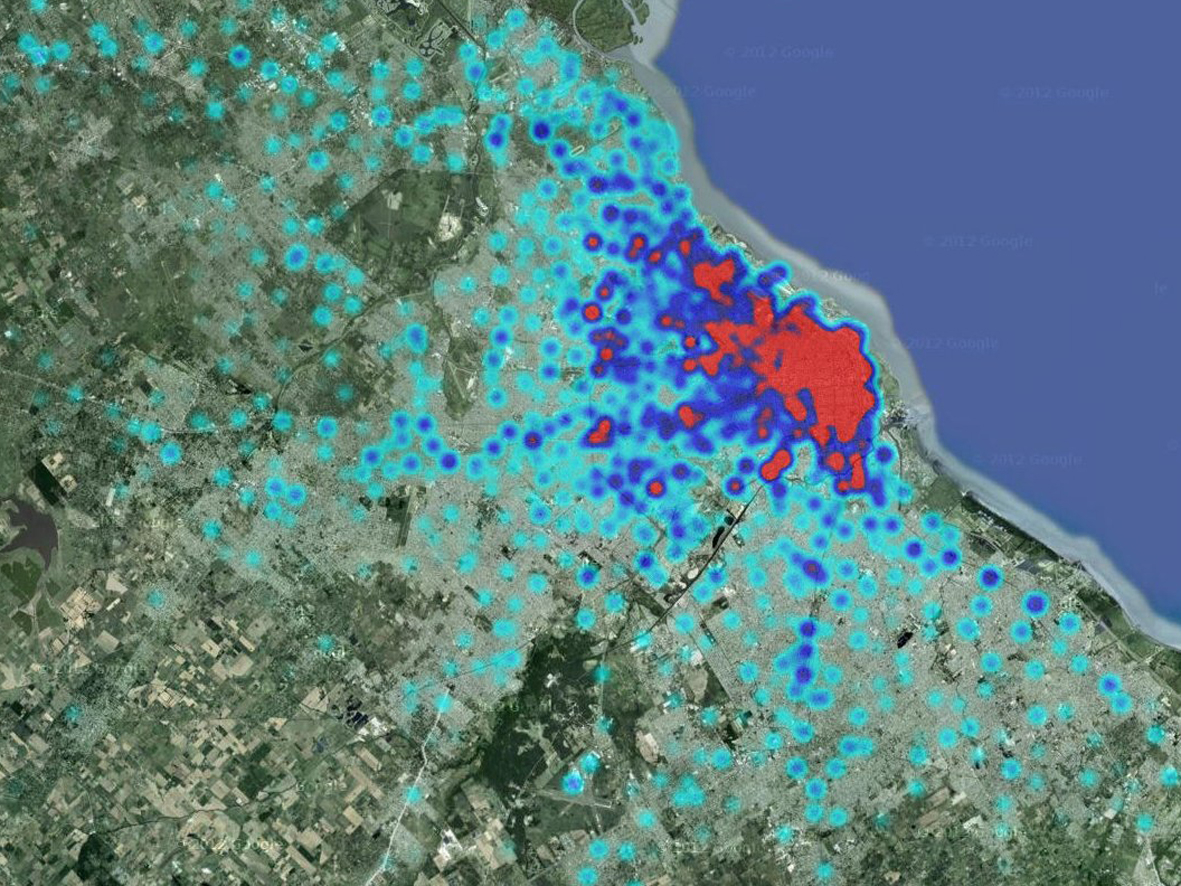}
  (d) 5 p.m.
\end{minipage}
%\hfill
\begin{minipage}{.33\textwidth}
  \centering
  \includegraphics[width=0.95\textwidth]{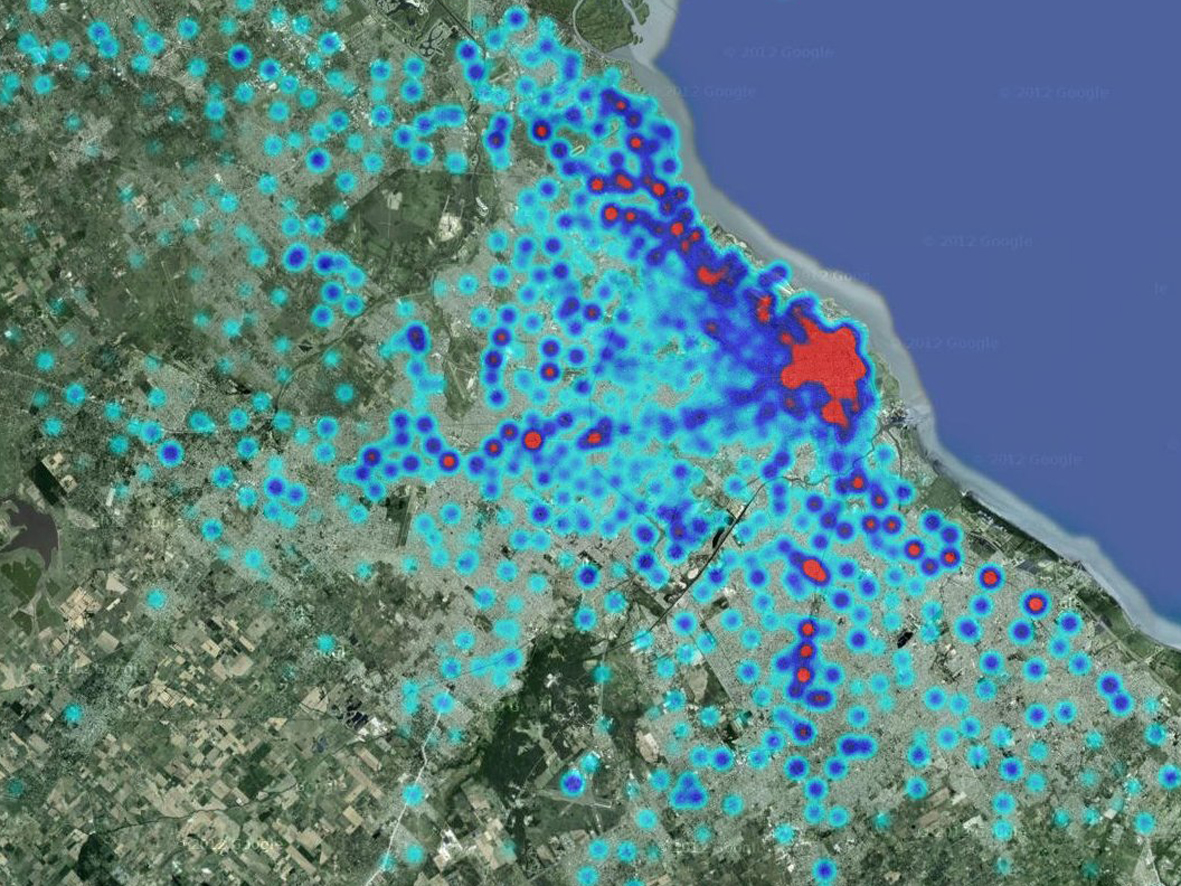}
  (e) 7 p.m.
\end{minipage}
\begin{minipage}{.33\textwidth}
  \centering
  \includegraphics[width=0.95\textwidth]{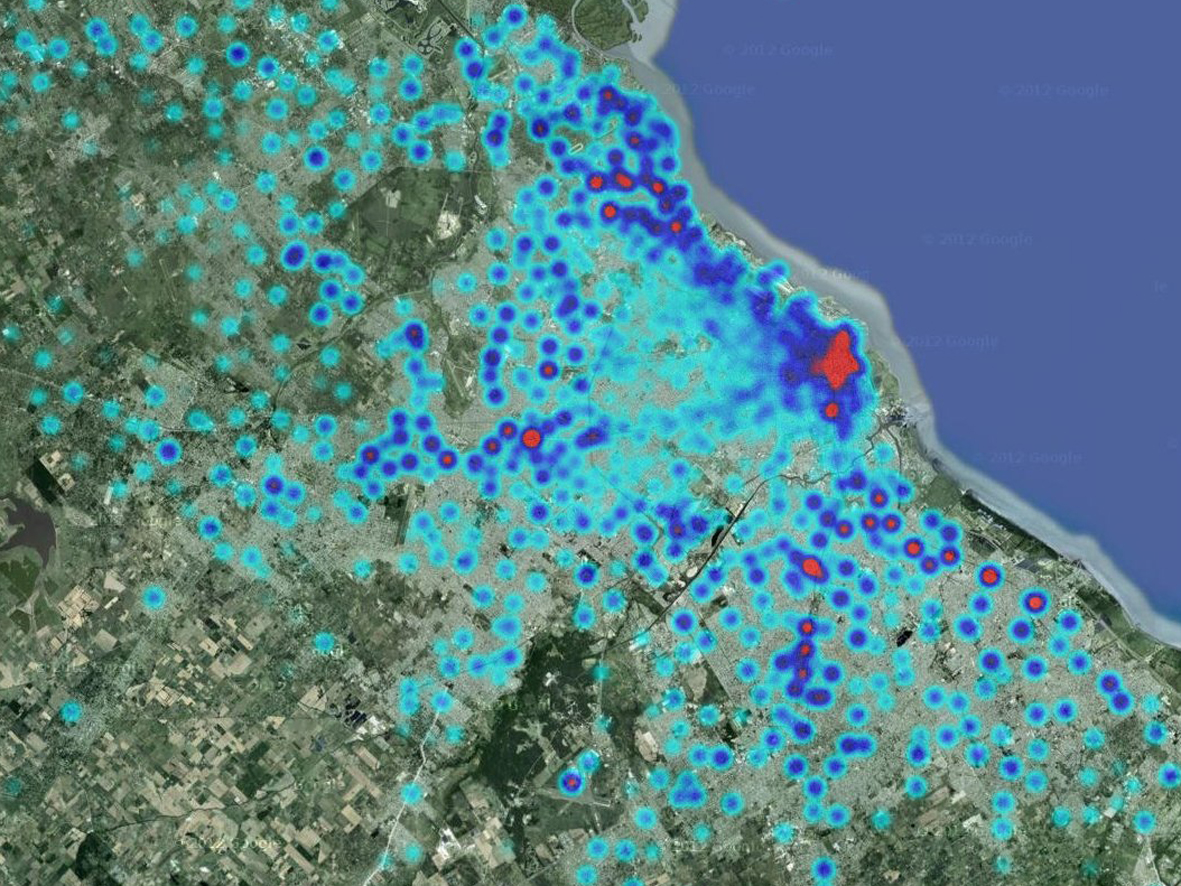}

  (f) 8 p.m.
\end{minipage}

  \caption{\label{fig:commute} Commute to a major city from the surrounding areas on a weekday,
   for different hours.
    Red color corresponds to a higher number of calls, whereas blue corresponds to an intermediate number of calls and light blue to a smaller one. 
    Reproduced from \cite{Ponieman2013human}.}
\vspace{-0.0cm}
\end{figure*}

The massive amounts of geolocation data collected from mobile phone records has sparked an ongoing effort to understand and predict the mobility patterns of human beings. In \cite{Ponieman2013human}, the authors study the extent to which social phenomena are reflected in mobile phone data, focusing in particular on the cases of urban commute and major sports events. The paper illustrates how these events are reflected in the data, and shows how information about the events can be used to improve predictability in a simple model for a mobile phone user's location.
As an illustration, Fig.~\ref{fig:commute} shows the commute to a major city from the surrounding areas on a weekday.
The authors of \cite{Ponieman2015mobility} further propose a method for the automatic detection of such events and discuss their relation to the social fabric as derived from mobile phone communications.

For urban and transportation planning, it is important to have tools to understand the mobility patterns of people in cities. An analysis of the mobility flows that occur in a city allows for appropriate infrastructure or policies decisions, that provide  transportation alternatives and incentivize efficient ways of moving, especially taking into account the problems of congestion and environmental pollution caused by transportation and its direct relationship with the health and quality of life of the population.

To analyze urban mobility, a series of methodologies are used that collect information on the trips of the population and allow city planners to understand how people move daily. These methodologies consist mainly of conducting household mobility surveys, travel journals or interception surveys in public and private transportation. These studies are methodologically complex, requiring long implementation times and a significant amount of economic resources. Although they provide extremely valuable information, most cities carry out these studies every five or ten years and other cities even more sporadically.

Currently, with the development of new technologies, large amounts of data are available (that would have been impossible to imagine years ago). This data revolution is impacting our way of conducting analysis and decision-making in different sectors, including in urban and transportation planning, generating new opportunities to analyze and better understand our cities.
The main objective of the study \cite{Anapolsky2014exploracion} is to discuss the methodological exploration carried out to analyze mobile phone data. This study discusses to what extent questions related to mobility in the territory can be answered, using this source of information.

Studying the activity of mobile phones allows us, not only to infer social networks between individuals, but also to observe the movements of these individuals in space and time. 
In \cite{Sarraute2015social}, the authors investigate how these two related sources of information can be integrated within the context of detecting and analyzing large social events. They show that large social events can be characterized not only by an anomalous increase in activity of the antennas in the neighborhood of the event, but also by an increase in social relationships of the attendants present in the event. Moreover, having detected a large social event via increased antenna activity, they make use of the network connections to infer whether an unobserved user was present at the event. More precisely, the paper addresses the following three challenges: (i) automatically detecting large social events via increased antenna activity; (ii) characterizing the social cohesion of the detected event; and (iii) analyzing the feasibility of inferring whether unobserved users were in the event.

\begin{figure*}[ht]
\vspace{-0.2cm}
\centering
\begin{minipage}{.42\linewidth}
  \centering
  \includegraphics[width=0.80\textwidth, clip=true, trim=58 40 10 20]{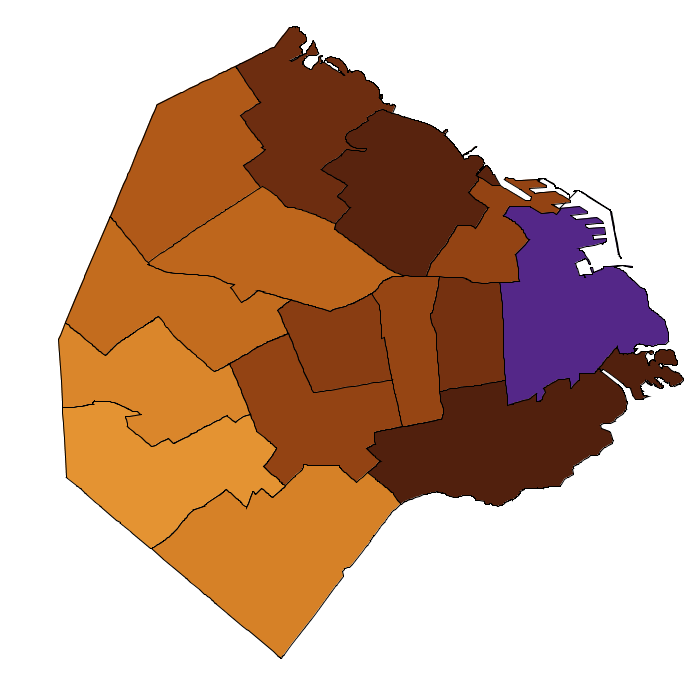}

  \small{(a)}
\end{minipage}
\begin{minipage}{.42\linewidth}
  \centering
  \includegraphics[width=0.80\textwidth, clip=true, trim=58 40 10 20]{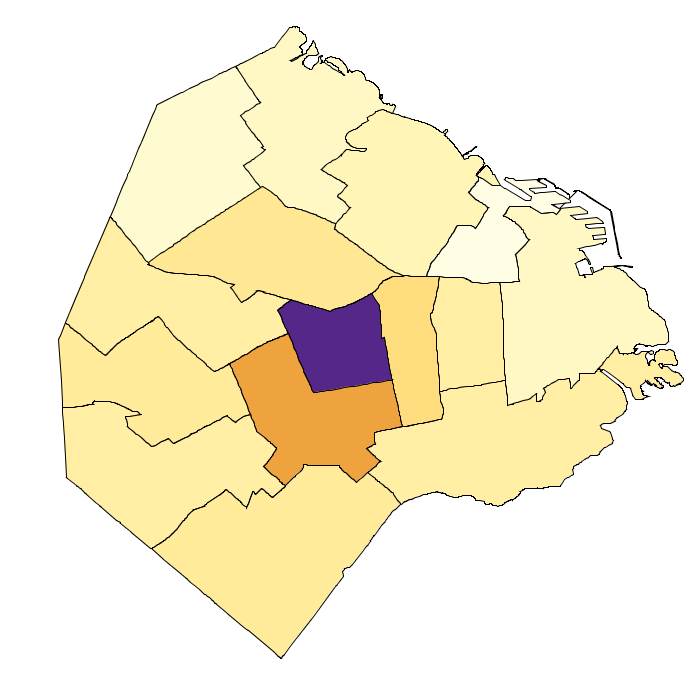}

  \small{(b)}
\end{minipage}
\begin{minipage}{.076\linewidth}
  \centering
  \includegraphics[width=0.85\textwidth]{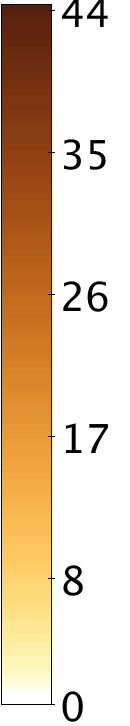}
  \phantom{\small{(c)}}
\end{minipage}
  \caption{\label{fig:visualization} Visualization of the number of people present on Monday to Thursday noon period in (a) Commune 1 and (b) Commune 6 (colored in violet) that live in each of the other communes of a major city. The scale shows the number of people (in thousands) each color represents. 
  Reproduced from \cite{Sarraute2015city}.}
\label{fig:city_pulse_vis}
\end{figure*}

Cell phone technology generates massive amounts of data. Although this data has been gathered for billing and logging purposes, today it has a much higher value, because its volume makes it very useful for big data analyses. In \cite{Sarraute2015city}, the authors analyze the viability of using cell phone records to lower the cost of urban and transportation planning, in particular, to find out how people travel in a specific city. They use anonymized cell phone data to estimate the distribution of the population in the city at different periods of time. They compare those results with traditional methods (urban polling) using data from origin-destination surveys. Traditional polling methods have a much smaller sample, in the order of tens of thousands (or even less for smaller cities), to maintain reasonable costs. Furthermore, these studies are performed at most once per decade, in the best cases, in many countries. 
The paper's objective is to prove that new methods based on cell phone data are reliable, and can be used indirectly to keep a real-time track of the flow of people among different parts of a city. It also goes further to explore new possibilities that have been opened by these methods. Fig.~\ref{fig:city_pulse_vis} shows a visualization of the number of people present in different communes of a city, according to their home commune.

Understanding human mobility patterns is crucial to fields such as urban mobility and mobile network planning. For this purpose, \cite{Mucelli2016regularity} makes use of large-scale datasets recording individuals' spatio-temporal locations, from eight major world cities: Beijing, Tokyo, New York, Paris, San Francisco, London, Moscow and Mexico City. Our contributions are two-fold: first, we show significant similarities in people’s mobility habits regardless of the city and nature of the dataset. Second, we unveil three persistent traits present in an individual’s urban mobility: repetitiveness, preference for shortest-paths, and confinement. These characteristics uncover people’s tendency to revisit few favorite venues using the shortest-path available.

%-----------------------------------------------------
\section{Mobility and Data Traffic Usage}
\label{data-traffic-usage}
%-----------------------------------------------------

\begin{figure}[ht]
\centering
\makebox[0pt]{\includegraphics[width=\linewidth]{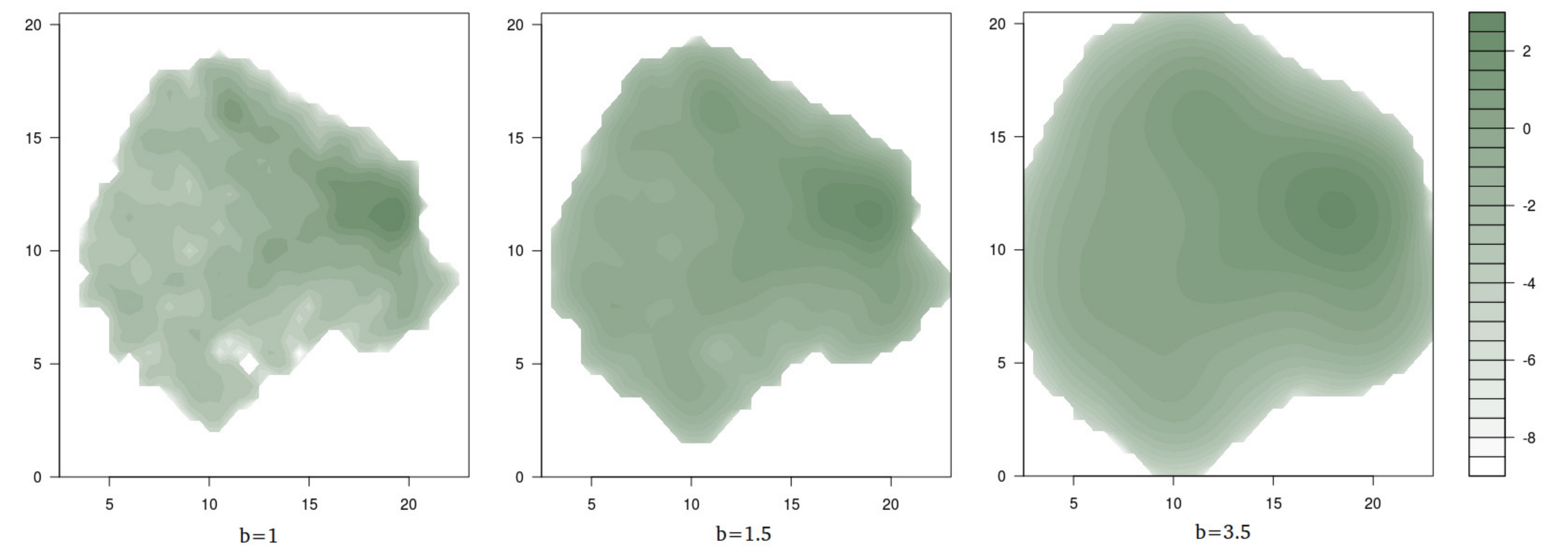}}
\caption{Contour plots of the intensity function in logaritmic scale,
showing the non homogeneous distribution of antennas in the city. 
Reproduced from \cite{Fattori2017new}.}
\label{fig:3intensidades}
\end{figure}

Paper \cite{Fattori2017new} introduces a model for the spatial distribution of cell towers in a major city. After showing that the Complete Spatial Randomness (homogeneous Poisson distribution) hypothesis does not hold, the authors propose a model in which each site is distributed according to a bivariate Gaussian variable with mean given by the barycenter of its neighbors in the Delaunay triangulation. They show that this model is suitable, and can be used to generate a synthetic distribution of cell towers.
Fig.~\ref{fig:3intensidades} shows the contour plots of intensity functions which the non homogeneous distribution of antennas in the city.

Using a large-scale dataset collected from a major 3G network in a dense metropolitan area, the paper \cite{Oliveira2015measurement} presents the first detailed measurement-driven model of mobile data traffic usage of smartphone subscribers. Their main contribution is a synthetic, measurement-based, mobile data traffic generator capable of simulating traffic-related activity patterns for different categories of subscribers and time periods for a typical day in their lives. The authors first characterize individual subscribers routinary behaviour, followed by a detailed investigation of subscribers' temporal usage patterns (i.e., ``when'' and ``how much'' traffic is generated). They classify the subscribers into six distinct profiles according to their usage patterns and model these profiles according to two daily time periods: peak and non-peak hours. The study shows that the synthetic trace generated by the data traffic model consistently replicates a subscriber's profile for these two time periods when compared to the original dataset. Broadly, their observations bring important insights into network resource usage. The authors also discuss relevant issues in traffic demands and describe implications in network planning and privacy.

The 3G cellular networks are struggling due to the pervasiveness of mobile data consumption.
The steady growth of smartphones, the very rapid evolution of services and their usage is accentuated in metropolitan scenarios due to the high urbanization and concentration of mobile users. In this context, understanding mobile data traffic demands per user is crucial for the evaluation of data offloading solutions designed to alleviate cellular networks.
The pervasive era also brought new facilities: currently smartphones provide the best means of gathering users information about content consumption behavior on a large scale. In this context, the literature is rich in work studying and modeling users mobility, but little is publicly known about users content consumption patterns.
The contributions of \cite{Mucelli2015analysis} are twofold: first, their analyses provide a precise characterization of individual subscribers traffic behavior clustered by their usage pattern, instead of a network-wide data traffic view. Second, they provide a traffic generator that synthetically, still consistently, reproduces real traffic demands. A synthetic traffic generator has positive implications on network resource allocation planning and testing, or hotspot deployment. Moreover, the synthetic traffic carries no direct personal information from the original users, thus greatly reducing privacy issues.

%-----------------------------------------------------
\section{Delay Tolerant Networks}
\label{delay-tolerant-networks}
%-----------------------------------------------------

In \cite{Leo2015taking}, from a measurement study and analysis of SMS based on traces coming from a nationwide cellular telecommunication operator during a two month period, the authors propose a DTN (Delay Tolerant Network) like network protocol for delivering SMS. More precisely, they perform a temporal and spatial analysis of a cellular network considering geolocalized SMS. The temporal analysis allows them to detect events and to check for overloading periods, with abnormal or unexpected traffic, and to study the evolution of classical parameters such as activity or distance between source and destination. The spatial analysis is based on the Voronoï diagram of the base stations covering a city. The paper explains how SMS traffic can be characterized. Such key characterization allows us to answer the question: is it possible to transmit SMS using phones as relay in a large city? They defined a simple network protocol to transmit SMS from a source to a destination. This DTN like protocol does not need routing nor global knowledge. The protocol takes benefit from the locality of SMS, the density of phones and the mobility of phone users. The protocol is studied with a mobile dataset including 8 millions users. This provided a precise estimation of the average transmission time and the global performance of this approach. After 30 minutes, half of the SMS were delivered successfully to their destination.

Cellular technologies are evolving quickly to constantly adapt to new usage and tolerate the load induced by the increasing number of phone applications. Understanding the mobile traffic is thus crucial to refine models and improve experiments. In this context, one has to understand the temporal activity of a user and the user's movements. At the user scale, the usage is not only defined by the number of calls but also by the user’s mobility. At a higher level, the base stations have a key role on the quality of service. In \cite{Leo2016call,Leo2016traces}, the authors analyze a very large set of Call Detail Records (CDR) over 12 months. It contains 8 million users and 5 billion call events. The first contribution is the study of call duration and inter-arrival time parameters. Then, they assess user movements between consecutive calls (switching from a station to another one). This study suggests that user mobility is pretty dependent on user activity. Furthermore, it shows properties of the inter-call mobility by making an analysis of the call distribution.

In urban areas, the population density is still growing (the population density starts exceeding 20.000 inhabitants per km2), and so, the density of mobile users becomes very important. People are moving from home to work, from work to active places. One can benefit from the mobility and the density to justify DTN (Delay Tolerant Network) approach protocol to convey SMS (or alternative messaging services) traffic. Indeed, the mobility of users, especially during the day, creates an ad hoc mobile network where the nodes are the smartphones held by mobile clients. In \cite{Leo2016performance}, their performance evaluations are based on a measurement and analysis of SMS traces coming from a nationwide cellular telecommunication operator during a two month period. The authors propose several DTN like basic network protocols for delivering SMS. They perform a temporal and spatial analysis of the cellular network in a major city considering geolocalized SMS to characterize the traffic. Such key characterization allows them to answer the question: is it possible to transmit SMS using phones as relay in a large city? Four network protocols are used to transmit SMS from a source to a destination. By leveraging a mobile dataset including 8 million users, they give a precise estimation of the average transmission time and the global performance of this approach. The analysis shows that after 30 min, half of the SMS are delivered successfully to their destination. Contrary to the cellular networks, the authors explain how much of the potentiality of the mobile users network can benefit from complementary properties such as the locality of SMS, the density of phones and the mobility of phone users. Moreover, they show that in a realistic scenario, this approach induces reasonable storage cost.

%-----------------------------------------------------
\section{Mobility and Epidemiology}
\label{epidemiology}
%-----------------------------------------------------

The study \cite{deMonasterio2016analyzing} uses mobile phone records for the analysis of mobility patterns and the detection of possible risk zones of Chagas disease in two Latin American countries. The authors show that geolocalized call records are rich in social and individual information, which can be used to infer whether an individual has lived in an endemic area. They present two case studies, in Argentina and in Mexico, using data provided by mobile phone companies from each country. The risk maps generated can be used by health campaign managers to target specific areas and allocate resources more effectively.

\begin{figure}[ht]
\begin{minipage}{.495\linewidth}
\centering
  \includegraphics[width=0.98\linewidth]
  {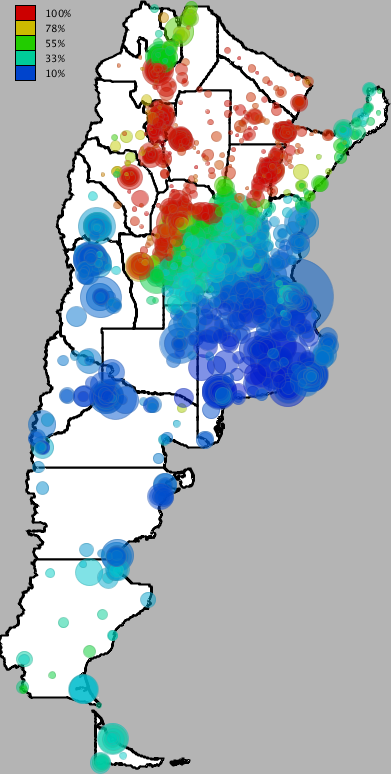}
  
(a) $\beta = 0.01$
\end{minipage}
\begin{minipage}{.495\linewidth}
\centering
  \includegraphics[width=0.98\linewidth]
  {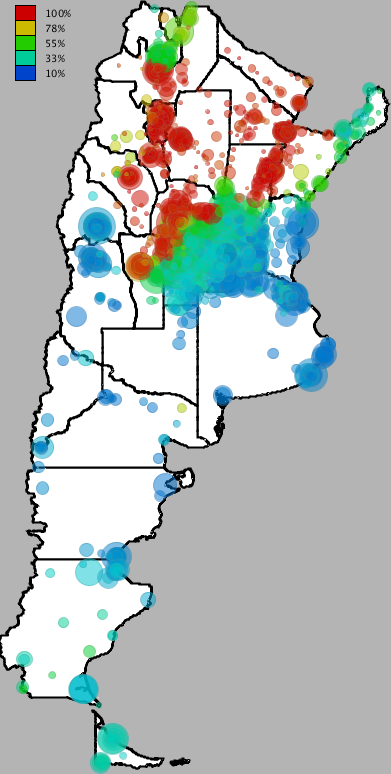}
  
(b) $\beta = 0.15$
\end{minipage}
\caption{Risk map for Chagas disease, filtered according to $\beta$ (minimum fraction of users having social ties to the endemic region $E_Z$). Reproduced from \cite{deMonasterio2016analyzing}.}
\label{fig:mapa_argentina}
\end{figure}

The heatmaps shown in Fig.~\ref{fig:mapa_argentina} expose an expected ``temperature'' descent from the endemic regions outwards. 
The authors also found communities atypical compared to their neighboring region, which stand out for their strong communication ties with the endemic region $E_Z$.
The detection of these communities is of great value to health campaign managers, providing them tools to target specific areas.
In summary, the paper~\cite{deMonasterio2016analyzing} shows the value of generating risk maps in order to prioritize effectively detection and treatment 
for the Chagas disease. 

The study \cite{Monasterio2017uncovering} additionally shows the value of mobile phone records to predict long-term migrations, which play a crucial role in the spread of Chagas disease.
In particular, the authors show that it is possible to use the mobile phone records of users during a bounded period of time in order to predict whether they have lived in the endemic zone $E_Z$ in a previous time frame.

%-----------------------------------------------------
\section{Predictability of Human Mobility}
\label{mobility-predictability}
%-----------------------------------------------------

The knowledge of the upper bounds of mobile data traffic predictors provides not only valuable insights into human behavior but also new opportunities to reshape mobile network management and services. It also provides researchers with insights into the design of effective prediction algorithms. In \cite{Chen2017spatio}, the authors leverage two large-scale real-world datasets collected by a major mobile carrier to investigate the limits of predictability of cellular data traffic demands generated by individual users. Using information theory tools, they measure the maximum predictability that any algorithm has the potential to achieve. They first focus on the predictability of mobile traffic consumption patterns in isolation. Their results show that it is theoretically possible to anticipate the individual demand with a typical accuracy of 85\% and reveal that this percentage is consistent across all user types. Despite the heterogeneity of users, they also find no significant variability in predictability when considering demographic factors or different mobility or mobile service usage. Then, the joint predictability of the traffic demands and mobility patterns is analyzed. They find that the two dimensions are correlated, which improves the predictability upper bound to 90\% on average.

Call Detail Records (CDRs) are a primary source of whereabouts in the study of multiple mobility-related aspects. However, the spatiotemporal sparsity of CDRs often limits their utility in terms of the dependability of results. Paper \cite{Chen2017towards}, driven by real-world data across a large population, proposes two approaches for completing CDRs adaptively, to reduce the sparsity and mitigate the problems the latter raises. Owing to high-precision sampling, the comparative evaluation shows that the authors' approaches outperform the legacy solution in the literature in terms of the combination of accuracy and temporal coverage. Also, they reveal important factors for completing sparse CDR data, which shed light on the design of similar approaches.

The literature is rich in mobility models that aim to predict human mobility. Yet, these models typically consider only a single kind of data source, such as data from mobile calls or location data obtained from GPS and web applications. Thus, the robustness and effectiveness of such data-driven models from the literature remain unknown when using heterogeneous types of data. In contrast, paper~\cite{Silveira2016mobhet} proposes a novel family of data-driven models, called MobHet, to predict human mobility using heterogeneous data sources. This proposal is designed to use a combination of features capturing the popularity of a region, the frequency of transitions between regions, and the contacts of a user, which can be extracted from data obtained from various sources, both separately and conjointly. The study evaluates the MobHet models, comparing them among themselves and with two single-source data-driven models, namely SMOOTH and Leap Graph, while considering different scenarios with single as well as multiple data sources. The experimental results show that the best MobHet model produces results that are better than or at least comparable to the best baseline in all considered scenarios, unlike the previous models whose performance is very dependent on the particular type of data used. These results thus attest to the robustness of the proposed solution for using heterogeneous data sources in predicting human mobility.